# Teleporting the one-qubit state via two-level atoms with spontaneous emission

Ming-Liang Hu[*]

*School of Science, Xi'an University of Posts and Telecommunications, Xi'an 710061, China*

**Abstract:** We study quantum teleportation via two two-level atoms coupled collectively to a multimode vacuum field and prepared initially in different atomic states. We concentrated on influence of the spontaneous emission, collective damping and dipole-dipole interaction of the atoms on fidelity dynamics of quantum teleportation and obtained the region of spatial distance between the two atoms over which the state can be teleported nonclassically. Moreover, we showed through concrete examples that entanglement of the channel state is the prerequisite but not the only essential quantity for predicting the teleportation fidelity.

## 1. Introduction

Quantum entanglement opens the door to many interesting applications of quantum mechanics in the field of information processing in the last decade. One of the most peculiar and fascinating one is quantum teleportation [1], by which the unknown state of a system can be transmitted from a location to a distant one with the help of local operations and classical communication (LOCC). The practical realization of teleportation relies crucially on the complete experimental control over a system's quantum state and the prior shared maximally entangled state resource between the sender (called Alice) and the receiver (called Bob). However, in a realistic scenario, entanglement is susceptible to local interactions with the ambient environment [2, 3], which can result in loss of coherence, and makes the quantum channel to be mixed and non-maximally entangled. In this case, the teleportation process generally becomes imperfect and Alice cannot teleport the unknown state to Bob with unit fidelity.

By virtue of the present atom trapping and cooling techniques, it is possible to trap two atoms at distances of the order of a resonant wavelength [4–6]. Based on these, theoretical generation of entanglement between atomic qubits has also been reported recently [7–9]. The results show that while the irreversible spontaneous decay of the atoms can lead to disentanglement of the initially entangled quantum states, by suitably controlling the interatomic separations it can also be used to create transient entanglement between two initially separable atoms [8, 9]. Since entanglement has always been considered as an essential physical resource for various tasks of quantum information processing, a question naturally arises at this stage is whether this entanglement can be used for quantum teleportation? If it is, how much information Alice can send to Bob? If it isn't, what is the underlying physical attribution behind it, or more explicitly, what ultimately determines the quality of the teleported state except amount of the entanglement of the resource? We will address these problems by concrete examples.

Since in real circumstances it is usually very difficult to prepare maximally entangled channel states and perform the unitary operations to the qubits, a number of authors investigated quantum teleportation with system decoherence [10–16] and noisy operations [10] in recent years, and the results showing that the environmental effects may cause the teleportation to lose its quantum advantage over purely classical communication [17]. Experimentally, quantum teleportation has also been demonstrated successfully with atomic qubits [18, 19]. In this paper, we investigate quantum teleportation of the one-qubit state with two spatially separated two-level atoms serving as the quantum channel. To see environmental effects on quality of teleportation, we assume that the two atoms coupled collectively to a multimode vacuum field. We restrict our concern to the influence of the spontaneous emission and the distances between the two atoms on fidelity (see section 2) of quantum teleportation with different initial atomic states, and compare the robustness between them. For the original proposal stated by Bennett *et al* [1], the system is considered to be isolated perfectly from its surrounding environment and the maximally entangled EPR pair is unitarily evolved, thus teleportation with unit fidelity can always be achieved. In the present case, however, the dissipative process of spontaneous emission may severely undermine the feasibility of the teleportation protocol.

---

[*] Corresponding author.
 E-mail: mingliang0301@xupt.edu.cn, mingliang0301@163.com (M.-L. Hu)

## 2. Basic formalism

The two identical two-level atoms serving as quantum channel for teleportation are assumed to be located at fixed positions $r_1$ and $r_2$ and connected by the dipole moments $\mu$. The lower and upper levels of them are denoted as $|g_i\rangle$ and $|e_i\rangle$ ($i = 1, 2$), which are separated by energy gap $\hbar\omega_0$, with $\omega_0$ being the transition frequency. Moreover, the two atoms are assumed to be coupled to all modes of the multimode vacuum electromagnetic field, under the influence of which the time evolution of the system is governed by the following master equation [20, 21]

$$\frac{\partial \rho}{dt} = -i\omega_0 \sum_{i=1}^{2}[S_i^z, \rho] - i\sum_{i \neq j}^{2} \Omega_{ij}[S_i^+ S_j^-, \rho] + \frac{1}{2}\sum_{i,j=1}^{2}\gamma_{ij}(2S_j^- \rho S_i^+ - S_i^+ S_j^- \rho - \rho S_i^+ S_j^-), \quad (1)$$

where $S_i^+$, $S_i^-$ and $S_i^z$ are the atomic dipole operators for the $i$th atom which satisfying the well-known commutation relations $[S_i^+, S_j^+] = 2S_i^z \delta_{ij}$ and $[S_i^z, S_j^\pm] = \pm S_i^\pm \delta_{ij}$, with $\delta_{ij}$ being the Kronecker delta. $\gamma_{ij} \equiv \gamma$ ($i = j$) are the spontaneous emission rates of the atoms caused by their direct interaction with the multimode vacuum field. Moreover, $\gamma_{ij}$ and $\Omega_{ij}$ ($i \neq j$) describe the collective damping and the dipole-dipole interaction potential, respectively. They both depend on the interatomic distance $r_{ij} = |r_j - r_i|$ and are defined as [20]

$$\gamma_{ij} = \frac{3}{2}\gamma\left\{[1-(\hat{\mu}\cdot\hat{r}_{ij})^2]\frac{\sin(kr_{ij})}{kr_{ij}} + [1-3(\hat{\mu}\cdot\hat{r}_{ij})^2]\left[\frac{\cos(kr_{ij})}{(k_0 r_{ij})^2} - \frac{\sin(kr_{ij})}{(kr_{ij})^3}\right]\right\},$$

$$\Omega_{ij} = \frac{3}{4}\gamma\left\{[(\hat{\mu}\cdot\hat{r}_{ij})^2 - 1]\frac{\cos(kr_{ij})}{kr_{ij}} + [1-3(\hat{\mu}\cdot\hat{r}_{ij})^2]\left[\frac{\sin(kr_{ij})}{(kr_{ij})^2} + \frac{\cos(kr_{ij})}{(kr_{ij})^3}\right]\right\}, \quad (2)$$

where $k = \omega_0/c = 2\pi/\lambda$ is the wave vector with $c$ and $\lambda$ being the velocity of light and the atomic resonant wavelength, respectively. $\hat{\mu}$ and $\hat{r}_{ij}$ are unit vectors along the transition dipole moment and along the interatomic axis. Moreover, we assume that the atomic dipole moments for the two atoms are parallel to each other, and in the following discussion we will first consider the case that they are polarized in the direction perpendicular to the interatomic axis (thus we have $\hat{\mu}\cdot\hat{r}_{ij} = 0$), and then extend it to general cases.

Without loss of generality, we assume that the one-qubit state Alice seeks to teleport to Bob is encoded at the atomic qubit A and can be expressed explicitly in the Bloch sphere representation as $|\varphi_{in}\rangle = \cos(\theta/2)|e\rangle + e^{i\phi}\sin(\theta/2)|g\rangle$, where $0 \leqslant \theta \leqslant \pi$ and $0 \leqslant \phi \leqslant 2\pi$ are the polar and azimuthal angles, respectively. Then the joint state composed of the state $\rho_{in} = |\varphi_{in}\rangle\langle\varphi_{in}|$ to be teleported and the quantum channel can be expressed as $\rho_{tot} = \rho_{in} \otimes \rho$. Alice is in possession of qubit A and the first qubit of the channel. To begin the teleportation process, she performs a Bell state measurement $\Pi_{A1}^k = |\Psi^k\rangle\langle\Psi^k|$ (with $|\Psi^{0,3}\rangle = (|e_1\rangle \otimes |e_2\rangle \pm |g_1\rangle \otimes |g_2\rangle)/\sqrt{2}$ and $|\Psi^{1,2}\rangle = (|e_1\rangle \otimes |g_2\rangle \pm |g_1\rangle \otimes |e_2\rangle)/\sqrt{2}$ being the four Bell states) on her two qubits, and then communicates classically the measurement result $k$ ($k = 0,1,2,3$) to Bob. After receiving two bits of classical information from Alice, Bob performs a conditional trace-preserving recovery operation $\mathcal{R}_m^{(k)}$ on qubit 2 to accomplish the teleportation process, and the final state is given by

$$\mathcal{E}_m^{(k)}[\rho] = \frac{1}{P_k} \mathcal{R}_m^{(k)}\{\text{tr}_{A,1}[(\Pi_{A1}^k \otimes \sigma_2^0)(\rho_{\text{in}} \otimes \rho)]\}, \tag{3}$$

where $\mathcal{R}_m^{(k)}\{\varrho\} = \sigma^m \varrho \sigma^m$, with the indexes $k$ and $m$ denoting the situation that Alice performs her measurement using the basis $\Pi_{A1}^k$ while Bob recovers the output state via the transformation $\sigma^m$. Here $\sigma^0$ and $\sigma^{1,2,3}$ denote the $2\times 2$ identity operator and the three Pauli spin operators, respectively. Moreover, $P_k = \text{tr}_{A,1,2}[(\Pi_{A1}^k \otimes \sigma_2^0)(\rho_{\text{in}} \otimes \rho)]$ is the probability for Alice to get the measurement outcome $k$.

**Table 1.** Dependence of $\chi_k^{(m)}$ on $\chi_m$ corresponds to different $\Pi_{A1}^k$ and $\sigma^m$.

| $km$ | $\chi_k^{(m)}$ | $km$ | $\chi_k^{(m)}$ | $km$ | $\chi_k^{(m)}$ | $km$ | $\chi_k^{(m)}$ |
|---|---|---|---|---|---|---|---|
| 00 | $\chi_0$ | 10 | $\chi_1$ | 20 | $\chi_2$ | 30 | $\chi_3$ |
| 01 | $\chi_1$ | 11 | $\chi_0$ | 21 | $\chi_3$ | 31 | $\chi_2$ |
| 02 | $\chi_2$ | 12 | $\chi_3$ | 22 | $\chi_0$ | 32 | $\chi_1$ |
| 03 | $\chi_3$ | 13 | $\chi_2$ | 23 | $\chi_1$ | 33 | $\chi_0$ |

The quality of the teleportation protocol in the present work can be evaluated by the average fidelity $\langle f[\rho(t)]\rangle$, which describes the fidelity $\langle \varphi_{\text{in}} | \mathcal{E}_{m_k}^{(k)}[\rho(t)] | \varphi_{\text{in}} \rangle$ averaged over all possible pure input states $|\varphi_{\text{in}}\rangle$ on the Bloch sphere and over all possible Alice's measurement outcomes $k$. In general, $\langle f[\rho(t)]\rangle$ can be written explicitly as follows

$$\langle f[\rho(t)]\rangle = \frac{1}{4\pi} \int_0^{2\pi} d\phi \int_0^\pi d\theta \sin\theta \sum_{k=0}^3 P_k \langle \varphi_{\text{in}} | \mathcal{E}_{m_k}^{(k)}[\rho(t)] | \varphi_{\text{in}} \rangle$$
$$= \sum_{k=0}^3 f_k^{(m_k)}, \tag{4}$$

where $4\pi$ is the solid angle, and here we use the notation $m_k$ ($m_k = 0,1,2,3$) instead of $m$ to signify different recovery operations performed by Bob. This is based on the consideration that for Alice's different measurement outcomes $k$, Bob may recover the state with different unitary transformations $\sigma^{m_k}$. Thus if we define the following four notations

$$\chi_{0,3} = \frac{1}{2}[\rho_{11}(t) + \rho_{44}(t) \pm \rho_{14}(t) \pm \rho_{41}(t)],$$
$$\chi_{1,2} = \frac{1}{2}[\rho_{22}(t) + \rho_{33}(t) \pm \rho_{23}(t) \pm \rho_{32}(t)]. \tag{5}$$

Then $f_k^{(m_k)}$ appeared in equation (4) can be derived analytically as $f_k^{(m_k)} = \chi_k^{(m_k)}/6 + 1/12$, this together with equation (4) yields

$$\langle f[\rho(t)]\rangle = \frac{1}{6} \sum_{k=0}^3 \chi_k^{(m_k)} + \frac{1}{3}, \tag{6}$$

where the explicit dependence of $\chi_k^{(m)}$ on $\chi_m$ with different $\Pi_{A1}^k$ and $\sigma^m$ are given in table 1. When Alice obtains the measurement result $k$, Bob can recover the state with any one of the four unitary transformations $\sigma^{m_k}$ ($m_k = 0,1,2,3$), but they give rise to different $\langle f[\rho(t)]\rangle$. For given channel state $\rho(t)$, one can maximize $\langle f[\rho(t)]\rangle$ by choosing proper series of $s = (m_0 m_1 m_2 m_3)$, or in other words, the maximum average fidelity $\langle f[\rho(t)]\rangle_{\max}$ achievable under Bob's appropriate recovery operations $\{\sigma^{m_0}\sigma^{m_1}\sigma^{m_2}\sigma^{m_3}\}$ is given by

$$F[\rho(t)] = \langle f[\rho(t)]\rangle_{\max} = \frac{2\mathcal{F}[\rho(t)]+1}{3}, \tag{7}$$

where we have used the notation $F[\rho(t)]$ to signify $\langle f[\rho(t)]\rangle_{\max}$ for the convenience of the following presentation, and $\mathcal{F}[\rho(t)]$ in Eq. (7) is in fact the fully entangled fraction [22] which can be expressed as

$$\mathcal{F}[\rho(t)] = \max_{k,m=0,1,2,3}\{\chi_k^{(m)}\} = \max_{n=0,1,2,3}\{\chi_n\}. \tag{8}$$

Moreover, in deriving the above equations, we have used the obvious fact that the magnitude of $\max_{m=0,1,2,3}\{\chi_k^{(m)}\}$ is independent of the values of $k$, which can be obtained directly from table 1.

## 3. Teleportation with two-level atoms

When the maximally entangled qubit pairs are used as quantum channel and they are protected ideally from decoherence, the relations between $k$ and $m$ are deterministic, i.e., for Alice's measurement result $k$, Bob's recovery operation $\sigma^m$ is unique and does not change with time [1]. For cases considered here, however, the channel state $\rho(t)$ established between Alice and Bob at an arbitrary time $t$ will be mixed due to the unavoidable coupling of the atoms to the radiation field. The explicit forms of $\rho(t)$ can be derived by solving the master equation (1) with $\rho(0)$ as the initial condition, but it is hard to obtain analytical solutions for general situations. Thus for simplicity, we consider in this work $\rho(0)$ in an 'X' formation [23], i.e., it only contains nonzero elements along the main diagonal and anti-diagonal. For this case, the nonzero elements of $\rho(t)$ can be obtained explicitly as

$$\begin{aligned}
\rho_{11}(t) &= \rho_{11}(0)e^{-2\gamma t}, \quad \rho_{14}(t) = \rho_{41}^*(t) = \rho_{14}(0)e^{-(\gamma+2i\omega_0)t}, \\
\rho_{22,33}(t) &= a_1\left[e^{-(\gamma+\gamma_{12})t} - e^{-2\gamma t}\right] + a_2\left[e^{-(\gamma-\gamma_{12})t} - e^{-2\gamma t}\right] \\
&\quad + b_1 e^{-(\gamma+\gamma_{12})t} + b_2 e^{-(\gamma-\gamma_{12})t} \pm c_1 e^{-(\gamma+2i\Omega_{12})t} \pm c_2 e^{-(\gamma-2i\Omega_{12})t}, \\
\rho_{23,32}(t) &= a_1\left[e^{-(\gamma+\gamma_{12})t} - e^{-2\gamma t}\right] - a_2\left[e^{-(\gamma-\gamma_{12})t} - e^{-2\gamma t}\right] \\
&\quad + b_1 e^{-(\gamma+\gamma_{12})t} - b_2 e^{-(\gamma-\gamma_{12})t} \pm c_1 e^{-(\gamma+2i\Omega_{12})t} \mp c_2 e^{-(\gamma-2i\Omega_{12})t}, \\
\rho_{44}(t) &= 1 - \rho_{11}(t) - \rho_{22}(t) - \rho_{33}(t),
\end{aligned} \tag{9}$$

where the corresponding parameters $a_{1,2}$, $b_{1,2}$ and $c_{1,2}$ are given by

$$\begin{aligned}
a_{1,2} &= \frac{1}{2}\rho_{11}(0)\frac{\gamma\pm\gamma_{12}}{\gamma\mp\gamma_{12}}, \\
b_{1,2} &= \frac{1}{4}[\rho_{22}(0)+\rho_{33}(0)\pm\rho_{23}(0)\pm\rho_{32}(0)], \\
c_{1,2} &= \frac{1}{4}[\rho_{22}(0)-\rho_{33}(0)\pm\rho_{23}(0)\mp\rho_{32}(0)].
\end{aligned} \tag{10}$$

In this work, we would like to explore in detail the relations between entanglement and purity of the channel state $\rho(t)$, and average fidelity of teleportation. Since the two atoms constitute the quantum channel are two-level systems, we quantify the amount of entanglement associated with $\rho(t)$ by using the concept of concurrence [24], which is unity for the maximally entangled states and zero for the separable (product) states. As stated by Wootters [24], the concurrence is defined as $C = \max\{0, \lambda_1 - \lambda_2 - \lambda_3 - \lambda_4\}$, where $\lambda_i$ ($i=1,2,3,4$) are the square roots of the eigenvalues of the operator $R = \rho(t)(\sigma^2 \otimes \sigma^2)\rho^*(t)(\sigma^2 \otimes \sigma^2)$ in decreasing order, and $\rho^*(t)$ denotes the complex conjugation of $\rho(t)$. Moreover, we measure purity of the two-atom system by the trace of the square of the density operator $\rho(t)$, i.e., $P = \text{tr}[\rho^2(t)]$.

Now we begin our discussions on fidelity dynamics of quantum teleportation of the one-qubit state with various types of concrete initial channel states $\rho(0)$. We will first consider the cases of $\rho(0)$ to be the four maximally entangled Bell states $|\Psi^k\rangle$ ($k = 0,1,2,3$), which all enable unit teleportation fidelity for the idealistic situation (i.e., no decoherence). Besides these, we would also like to explore behaviors of average teleportation fidelity with the two atoms of the channel prepared initially in a separable state $|e_1\rangle \otimes |g_2\rangle$. This consideration is stimulated by the fact that from this initial state one can entangling the two atoms with considerable amount of entanglement via the dissipative process of spontaneous emission [7,8], thus from the perspective of utility, it is natural for us to conjecture whether this entanglement is useful for teleportation.

We first see fidelity dynamics of quantum teleportation with the two atoms prepared initially in the symmetric state $\rho(0) = |\Psi^1\rangle\langle\Psi^1|$, from which one obtain $a_{1,2} = b_2 = c_{1,2} = 0$ and $b_1 = 1/2$. This gives rise to $\rho(t) = e^{-(\gamma+\gamma_{12})t}|\Psi^1\rangle\langle\Psi^1| + [1 - e^{-(\gamma+\gamma_{12})t}]|g_1g_2\rangle\langle g_1g_2|$ (here the notation $|g_1g_2\rangle = |g_1\rangle \otimes |g_2\rangle$), substituting of which into Eq. (5) one can obtain

$$\chi_{0,3} = \frac{1}{2}[1 - e^{-(\gamma+\gamma_{12})t}], \quad \chi_1 = e^{-(\gamma+\gamma_{12})t}, \quad \chi_2 = 0. \tag{11}$$

From the above equations one can see that the relative magnitudes of $\chi_m$ ($m = 0,1,2,3$) are dependent on the evolution time $t$. When $(\gamma+\gamma_{12})t < \ln 3$, we obtain the fully entangled fraction as $\mathcal{F}[\rho(t)] = \max_{m=0,1,2,3}\{\chi_m\} = \chi_1$. During this time region, the one-to-one correspondence between Alice's measurement basis and Bob's recovery operation should be $\Pi_{A1}^{0,1,2,3} \mapsto \sigma^{1,0,3,2}$ (see table 1) for the purpose of achieving the maximal average fidelity. When $(\gamma+\gamma_{12})t > \ln 3$, however, we have $\mathcal{F}[\rho(t)] = \max_{m=0,1,2,3}\{\chi_m\} = \chi_{0,3}$, and now the one-to-one correspondence between Alice's measurement basis and Bob's recovery operation should be $\Pi_{A1}^{0,1,2,3} \mapsto \sigma^{0,1,2,3}$ or $\Pi_{A1}^{0,1,2,3} \mapsto \sigma^{3,2,1,0}$. The origin of two different one-to-one correspondences in this time region can be understood from equation (3). For example, if Alice performs her measurement using the Bell basis $\Pi_{A1}^0$, then it is straightforward to check that Bob's transformations $\sigma^{0,3}$ yield the final output state with identical diagonal elements, while the anti-diagonal elements only differs through a phase factor $e^{i\pi}$, which clearly gives rise to $f_0^{(0)} = f_0^{(3)}$. Thus for Alice's measurement basis $\Pi_{A1}^0$, Bob can perform either transformation $\sigma^0$ or $\sigma^3$ to recover the state for achieving the maximal fidelity. The same analysis can also be applied to the remaining measurement basis $\Pi_{A1}^{1,2,3}$ of Alice as well as similar phenomena appeared in the following discussions with other initial atomic states.

It follows from Eqs. (7), (8) and (11) that

$$F[\rho(t)] = \begin{cases} \dfrac{2e^{-(\gamma+\gamma_{12})t} + 1}{3} & (t < \ln 3/(\gamma+\gamma_{12})), \\ \dfrac{2 - e^{-(\gamma+\gamma_{12})t}}{3} & (t > \ln 3/(\gamma+\gamma_{12})). \end{cases} \tag{12}$$

Moreover, the concurrence and the purity of the two-atom system associated with $\rho(t)$ can also be obtained analytically as

$$C = e^{-(\gamma+\gamma_{12})t}, \quad P = 1 - 2e^{-(\gamma+\gamma_{12})t} + 2e^{-2(\gamma+\gamma_{12})t}. \tag{13}$$

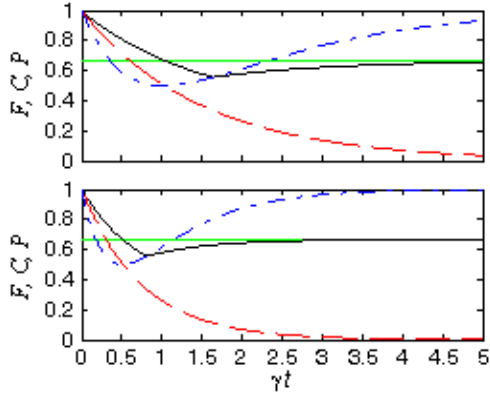

**Figure 1.** Average fidelity of teleportation (solid black curves), concurrence (dashed red curves) and purity (dash-dot blue curves) of the atomic state $\rho(t)$ versus the rescaled time $\gamma t$ with $\rho(0) =|\Psi^1\rangle\langle\Psi^1|$ (the top panel) and $\rho(0) =|\Psi^2\rangle\langle\Psi^2|$ (the bottom panel), respectively. Here $\hat{\mu} \perp \hat{r}_{12}$, and the interatomic separation is chosen to be $r_{12} = 0.6737\lambda$, at which the collective damping rate $\gamma_{12}$ attains its minimum. Moreover, the horizontal lines at $F = 2/3$ show the highest fidelity for classical transmission of a quantum state.

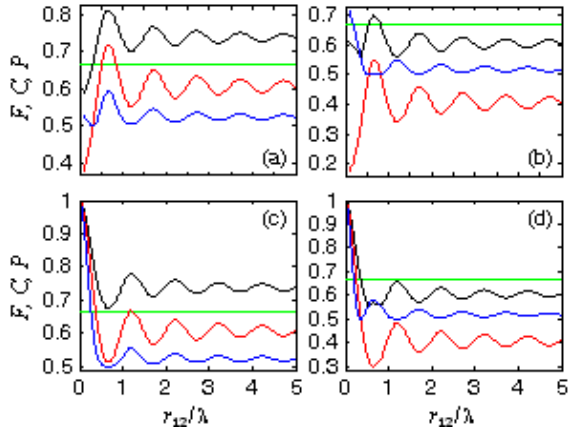

**Figure 2.** Average fidelity of teleportation (black curves), concurrence (red curves) and purity (blue curves) of the atomic state $\rho(t)$ versus $r_{12}/\lambda$ with $\rho(0) =|\Psi^1\rangle\langle\Psi^1|$ (the top two panels) and $\rho(0) =|\Psi^2\rangle\langle\Psi^2|$ (the bottom two panels), respectively. Here $\hat{\mu} \perp \hat{r}_{12}$, and the horizontal lines at $F = 2/3$ show the highest fidelity for classical transmission of a quantum state. Moreover, (a) and (c) are plotted with $\gamma t = 0.5$, while (b) and (d) are plotted with $\gamma t = 0.9$.

Plots of the above equations are shown in the top panel of figure 1 as functions of the rescaled time $\gamma t$ with fixed interatomic separations and in the top two panels of figure 2 as functions of the interatomic separations $r_{12}/\lambda$ with two different rescaled decay times. Figure 1 is plotted with $r_{12} = 0.6737\lambda$, at which the collective damping rate of the system attains its minimum value $\gamma_{12}^{\min} \simeq -0.3355\gamma$. During the region of $(\gamma+\gamma_{12})t < \ln 3$, the average fidelity $F[\rho(t)]$ decays monotonously with increasing rescaled time $\gamma t$ and becomes smaller than $2/3$ when $t$ increases after $t > t_c = \ln 2/(\gamma+\gamma_{12})$. $t_c$ behaves as damped oscillations with $r_{12}$ and attains its maximum of about $\ln 2/0.6645\gamma$ when the two atoms are separated by distance $r_{12} = 0.6737\lambda$. For fixed $\gamma t$, from Eq. (12) one can observe that $F[\rho(t)]$ increases with decreasing value of $\gamma_{12}$, thus figure 1 shows in fact the maximal average fidelity achievable via controlling the distance between the two atoms. In the region of $(\gamma+\gamma_{12})t > \ln 3$, however, $\chi_0$ or $\chi_3$ becomes the fully entangled fraction, and $F[\rho(t)]$ begins to increase with increasing $\gamma t$. But no matter how one adjusts the interatomic separations $r_{12}$, $F[\rho(t)]$ cannot exceed the classical limiting value of $2/3$ [17] due to the obvious fact that the exponential term in Eq. (12) is always larger than zero. Another interesting phenomenon which can be seen from figure 1 is that in the region of $(\gamma+\gamma_{12})t > \ln 3$, the increase of $F[\rho(t)]$ with time $\gamma t$ is accompanied by the decrease of the concurrence, which indicates that the larger amount of entanglement may not always ensure the higher teleportation fidelity. This is in contrast with the conclusion obtained in Refs. [10,11], in which the authors argued that the entanglement is a genuine resource for teleportation even if noises are involved, however, our results presented in figure 1 revealed that their statement is not universal. In fact, entanglement of the channel state is the prerequisite but not the only essential quantity for predicting fidelity of quantum teleportation [16,25]. This finding will be further confirmed in the following discussions with other types of initial atomic states.

For fixed decay time $\gamma t$, the average fidelity $F[\rho(t)]$ behaves as damped oscillations with the increase of the interatomic separations $r_{12}$. Two exemplified plots with $\gamma t=0.5$ and $\gamma t=0.9$ are presented in figures 2(a) and (b), respectively. For very small interatomic separations $kr_{12}$ goes to zero, and thus $\gamma_{12}$ reduces to $\gamma$ (corresponding to the small sample model or Dicke model [25]), for which $F[\rho(t)]$ attains the value of $(2e^{-2\gamma t}+1)/3$. Increasing the interatomic separations may enhance the magnitudes of $F[\rho(t)]$, and when $r_{12}=0.6737\lambda$ it attains a certain maximum value. Moreover, from Eqs. (2) and (12) one can obtain numerically that for short decay time $\gamma t < 0.5804$ (an exemplified plot is shown in figure 2(a) with $\gamma t=0.5$), one can teleport the one-qubit state with average fidelity better than that achievable via classical communication alone even if the two atoms are separated by infinite distances. Prolonging the decay time ($0.5804 < \gamma t < 1.0431$) will reduce the region of the interatomic separations $r_{12}$ for achieving nonclassical fidelity (see the exemplified plot presented in figure 2(b) with $\gamma t=0.9$), and when $\gamma t > 1.0431$ the teleportation protocol will failed completely.

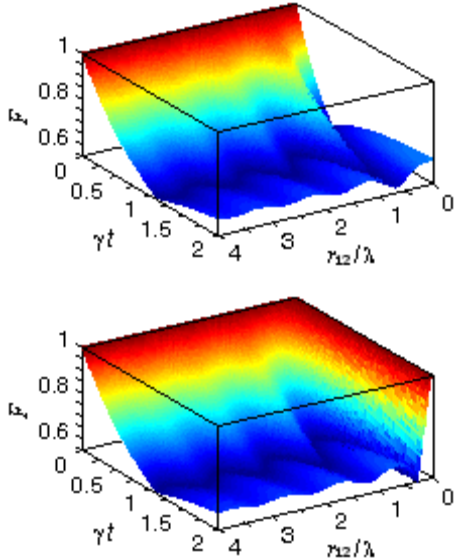

**Figure 3.** Average fidelity of teleportation versus $\gamma t$ and $r_{12}/\lambda$ with $\hat{\mu} \perp \hat{r}_{12}$, where the top and the bottom panels correspond to $\rho(0)=|\Psi^1\rangle\langle\Psi^1|$ and $\rho(0)=|\Psi^2\rangle\langle\Psi^2|$, respectively.

To show more intuitively the influence of the spontaneous emission and collective damping of the system on quantum teleportation, we display in the top panel of figure 3 the average fidelity versus the rescaled time $\gamma t$ and interatomic distance $r_{12}/\lambda$ with $\rho(0)=|\Psi^1\rangle\langle\Psi^1|$, from which one can also see that in the short-time region, the fidelity $F[\rho(t)]$ can be enhanced by adjusting the interatomic distances $r_{12}$, while in the long-time region, $F[\rho(t)]$ cannot exceeds the classical limiting value of $2/3$, no matter how one adjusts the interatomic distances.

Next we see fidelity dynamics of quantum teleportation with the two atomic qubits prepared initially in the antisymmetric state $\rho(0) = |\Psi^2\rangle\langle\Psi^2|$, for which we have $a_{1,2} = b_1 = c_{1,2} = 0$ and $b_2 = 1/2$. Combination of these with equation (9) we obtain the explicit form of $\rho(t)$ analytically as $\rho(t) = e^{-(\gamma-\gamma_{12})t}|\Psi^2\rangle\langle\Psi^2| + [1-e^{-(\gamma-\gamma_{12})t}]|g_1 g_2\rangle\langle g_1 g_2|$. This together with Eq. (5) yields

$$\chi_{0,3} = \frac{1}{2}[1 - e^{-(\gamma-\gamma_{12})t}], \ \chi_1 = 0, \ \chi_2 = e^{-(\gamma-\gamma_{12})t}. \tag{14}$$

Similar to the situation of $\rho(0) = |\Psi^1\rangle\langle\Psi^1|$, the maximum of $\chi_m$ ($m = 0,1,2,3$) is still time dependent. In the region of $(\gamma - \gamma_{12})t < \ln 3$, we have $\mathcal{F}[\rho(t)] = \max_{m=0,1,2,3}\{\chi_m\} = \chi_2$, and thus the one-to-one correspondence between Alice's measurement basis and Bob's recovery operation should be $\Pi_{A1}^{0,1,2,3} \mapsto \sigma^{2,3,0,1}$ for achieving the maximum teleportation fidelity. In the region of $(\gamma - \gamma_{12})t > \ln 3$, we have $\mathcal{F}[\rho(t)] = \max_{m=0,1,2,3}\{\chi_m\} = \chi_{0,3}$, and the one-to-one correspondence between Alice's measurement basis and Bob's recovery operation becomes $\Pi_{A1}^{0,1,2,3} \mapsto \sigma^{0,1,2,3}$ or $\Pi_{A1}^{0,1,2,3} \mapsto \sigma^{3,2,1,0}$. Thus the average fidelity in the whole time region takes the form

$$F[\rho(t)] = \begin{cases} \dfrac{2e^{-(\gamma-\gamma_{12})t} + 1}{3} & (t < \ln 3/(\gamma-\gamma_{12})), \\ \dfrac{2 - e^{-(\gamma-\gamma_{12})t}}{3} & (t > \ln 3/(\gamma-\gamma_{12})). \end{cases} \tag{15}$$

The concurrence and the purity of the channel state at an arbitrary time $t$ can also be obtained analytically, which are given by

$$C = e^{-(\gamma-\gamma_{12})t}, \ \ P = 1 - 2e^{-(\gamma-\gamma_{12})t} + 2e^{-2(\gamma-\gamma_{12})t}. \tag{16}$$

In the bottom panel of figure 1 we presented dynamical behaviors of the average fidelity, the concurrence and the purity of $\rho(t)$ with fixed interatomic separations $r_{12} = 0.6737\lambda$. Clearly, the plots show qualitatively similar behaviors with the former case. But now the critical time after which the teleportation protocol loses its quantum advantage over purely classical communication becomes $t_c = \ln 2/(\gamma - \gamma_{12})$. For very small interatomic separations, $t_c$ becomes very large and thus both $F[\rho(t)]$ and $C$ decay slowly. When the spatial distribution of the two atoms increases, $t_c$ first decreases to a certain minimum of about $\ln 2/1.3355\gamma$, and then behaves as damped oscillations with the same frequency describing the collective damping. Moreover, as can be obtained from Eq. (15), in the region of $(\gamma - \gamma_{12})t < \ln 3$, the average fidelity $F[\rho(t)]$ increases as $\gamma_{12}$ increases, thus the plots shown in the bottom panel of figure 1 is the minimum of $F[\rho(t)]$ achievable via adjusting the distances between the two atoms. This behaviors are also displayed in the bottom two panels of figure 2, from which one can see that for infinitesimal interatomic separations (i.e., $r_{12} \to 0$ and thus $\gamma_{12} \to \gamma$), the average fidelity $F[\rho(t)]$ is very close to unity. Increasing the spatial distances between the two atoms will diminish the values of both $F[\rho(t)]$ and $C$, and make them oscillating with increasing $r_{12}$, with the same frequency as $\gamma_{12}$. Moreover, as can be observed for figures 2(c) and (d), with short decay times ($\gamma t < 0.5190$) the one-qubit state can be teleported with nonclassical fidelity over arbitrary distances ($0 < r_{12} < \infty$), while for long decay times ($\gamma t > 0.5190$) it can only be teleported nonclassically over finite distances.

A more intuitive three dimensional figure of the average fidelity $F[\rho(t)]$ versus the rescaled time $\gamma t$ and interatomic separations $r_{12}/\lambda$ with $\rho(0) = |\Psi^2\rangle\langle\Psi^2|$ are displayed in the bottom panel of figure 3. It shows clearly that $F[\rho(t)]$ always attains its maximum unity for infinitesimal

interatomic separations, but its decay rate increases with increasing interatomic separations.

From the above analysis, one can see that for the initial atomic states $\rho(0)=|\Psi^1\rangle\langle\Psi^1|$ and $\rho(0)=|\Psi^2\rangle\langle\Psi^2|$, the spatial distances between the two atoms can significantly influence the teleportation fidelity. Now we turn our attention to the cases of $\rho(0)=|\Psi^{0,3}\rangle\langle\Psi^{0,3}|$. For these two types of initial states, from equations (9) and (10) it is straightforward to check that $\rho(t)$ have completely the same form except that $\rho_{14}(t)=\rho_{41}^*(t)=e^{-(\gamma+2i\omega_0)t}/2$ for $\rho(0)=|\Psi^0\rangle\langle\Psi^0|$, while $\rho_{14}(t)=\rho_{41}^*(t)=-e^{-(\gamma+2i\omega_0)t}/2$ for $\rho(0)=|\Psi^3\rangle\langle\Psi^3|$. By combination of these with Eq. (5) one can obtain that they yield the same average fidelity. Thus in the following discussion we only consider the case of $\rho(0)=|\Psi^0\rangle\langle\Psi^0|$. The four notations defined in Eq. (5) can be derived explicitly, however, we do not list them here for the calculation is direct and their expressions are rather lengthy. Instead, we plot in figure 4 the dynamical behaviours of the average fidelity $F[\rho(t)]$, as well as the concurrence $C$ and the purity $P$ of the atomic state $\rho(t)$ with different interatomic separations.

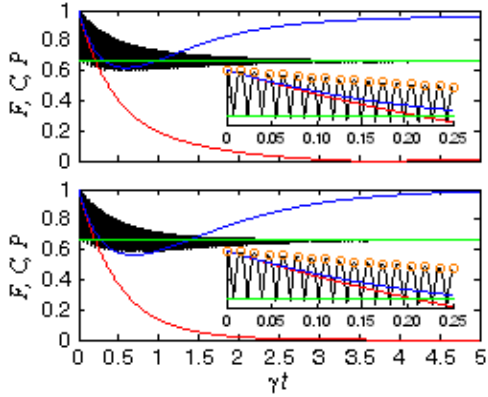

**Figure 4.** Average fidelity of teleportation (black curves), concurrence (red curves) and purity (blue curves) of the atomic state $\rho(t)$ versus the rescaled time $\gamma t$ with $\rho(0)=|\Psi^0\rangle\langle\Psi^0|$ and $\hat{\mu}\perp\hat{r}_{12}$, where the transition frequency of the two atoms has been chosen to be $\omega_0=100\gamma$. The other parameters for the plots are $r_{12}=\lambda/6$ (the top panel) and $r_{12}=\lambda/2$ (the bottom panel). Moreover, the hollow circles in the insets show time dependence of $F[\rho(t)]$ when $t=k\pi/2\omega_0$ ($k\in\mathbb{Z}$), and the horizontal lines at $F=2/3$ show the highest fidelity for classical transmission of a quantum state.

From equation (5) one can conclude that only $\chi_0$ or $\chi_3$ may contribute to the average fidelity $F[\rho(t)]$. Since they contain the terms $e^{-\gamma t}\cos(2\omega_0 t)$ and $-e^{-\gamma t}\cos(2\omega_0 t)$, respectively, the fully entangled fraction $\mathcal{F}[\rho(t)]$ becomes $\chi_0$ and $\chi_3$ alternatively with increasing decay time $\gamma t$. Thus in order to achieve the maximum average fidelity, the one-to-one correspondence between Alice's measurement basis and Bob's recovery transformation should be $\Pi_{A1}^{0,1,2,3}\mapsto\sigma^{0,1,2,3}$ when $\cos(2\omega_0 t)>0$, and $\Pi_{A1}^{0,1,2,3}\mapsto\sigma^{3,2,1,0}$ when $\cos(2\omega_0 t)<0$. Also due to the presence of the cosine terms $\cos(2\omega_0 t)$ in which the transition frequency $\omega_0\gg\gamma$, the average fidelity $F[\rho(t)]$ behaves as rapid oscillations with increasing decay time $\gamma t$, with frequency given by $4\omega_0$ (see figure 4). When $t=k\pi/2\omega_0$ ($k\in\mathbb{Z}$), the average fidelity $F[\rho(t)]$ attains its peak values. Moreover, as is evident from figure 4, we see that while both the entanglement and the purity of the atomic state $\rho(t)$ decrease, the average fidelity $F[\rho(t)]$ may be increased. Yeo has pointed out in a recent work [25] that the amount of mixing between the separable and maximally entangled states is also important in determining the average fidelity. For his model, the greater proportion of separable $\rho_{44}$ than the maximally entangled $|\Psi^2\rangle\langle\Psi^2|$ contributes to the poor quality of teleportation. For model considered here, however, increasing the proportions of separable $\rho_{11}(t)$ and $\rho_{44}(t)$ may increase the average fidelity $F[\rho(t)]$. We do not know what is the underlying physical attribution for this seemingly counterintuitive phenomenon. Does there exist other essential quantities except entanglement and purity of the channel state for predicting the teleportation fidelity? Further study is needed for clarifying this ambiguity. Mathematically, one could attribute the occurrence of this phenomenon to the presence of the phase factors $e^{\pm 2i\omega_0 t}$ appearing in $\rho_{41,14}(t)$, which contribute to the average fidelity $F[\rho(t)]$ but do not contribute to both the concurrence $C = \max\{0, C_1, C_2\}$ and the purity $P = \sum_{ij}|\rho_{ij}(t)|^2$, where $C_1 = 2[|\rho_{14}(t)| - \rho_{22}(t)]$ and $C_2 = 2[|\rho_{23}(t)| - \rho_{11}(t)]$.

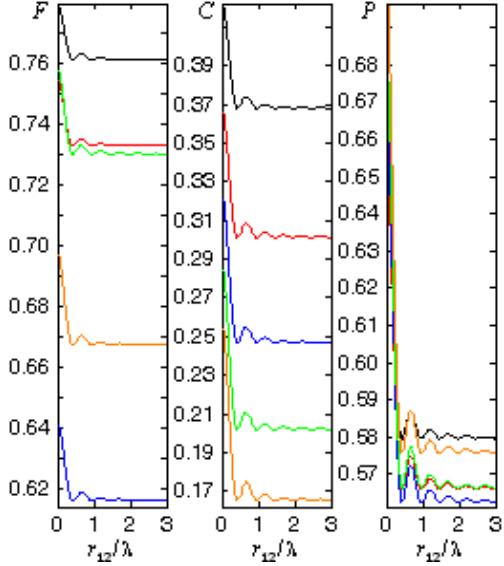

**Figure 5.** Average fidelity of teleportation, concurrence and purity of the atomic state $\rho(t)$ versus $r_{12}/\lambda$ with $\rho(0) = |\Psi^0\rangle\langle\Psi^0|$, $\omega_0 = 100\gamma$ and $\hat{\mu} \perp \hat{r}_{12}$. The other parameters for these plots are $\gamma t = 0.5$ (black curves), $\gamma t = 0.6$ (red curves), $\gamma t = 0.7$ (blue curves), $\gamma t = 0.8$ (green curves) and $\gamma t = 0.9$ (yellow curves).

Plots of the average fidelity $F[\rho(t)]$ as well as the concurrence $C$ and the purity $P$ of the atomic state $\rho(t)$ versus the spatial distances between the two atoms with different decay times were shown in figure 5. For very small interatomic separations $r_{12}$, the collective damping rate $\gamma_{12} \rightarrow \gamma$ (i.e., the small sample model or Dicke model [26]), from which one can derive the fully entangled fraction explicitly as $\mathcal{F}[\rho(t)] = [1 - \gamma t e^{-2\gamma t} + e^{-\gamma t}|\cos(2\omega_0 t)|]/2$, thus the average fidelity $F[\rho(t)]$ attains its maximum $F[\rho(t)] = [2 - \gamma t e^{-2\gamma t} + e^{-\gamma t}|\cos(2\omega_0 t)|]/3$. Similarly, in the limit of $\gamma_{12} \rightarrow \gamma$ the concurrence and the purity of the atomic state $\rho(t)$ also attain their maxima $C = e^{-\gamma t} - \gamma t e^{-2\gamma t}$ and $P = 1 - (2\gamma t + 0.5)e^{-2\gamma t} + (2\gamma^2 t^2 + \gamma t + 0.5)e^{-4\gamma t}$. All of these were evidently shown in figure 5. Also from figure 5 one can see that increasing the interatomic separations will diminish the values of $F[\rho(t)]$, $C$ and $P$, and makes them oscillating weakly with increasing $r_{12}$. For very large interatomic separations the collective damping rate $\gamma_{12}$ goes to zero (i.e., there is no coupling between the two atoms), and one can obtain $F[\rho(t)] = [2 + e^{-2\gamma t} + e^{-\gamma t}(|\cos(2\omega_0 t)| - 1)]/3$, $C = e^{-2\gamma t}$ and $P = 1 + e^{-4\gamma t} - 2e^{-3\gamma t} + 3e^{-2\gamma t} - 2e^{-\gamma t}$. Moreover, as is evident from figure 5, the state can be teleported with $F[\rho(t)] > 2/3$ over arbitrary long distances for certain time intervals, while for some other time intervals the state can only be teleported with $F[\rho(t)] > 2/3$ over finite ranges of distance or cannot be teleported nonclassically at all. Comparing the three plots in figure 5, one can also note that entanglement is not the only essential quantity for predicting the fidelity.

Finally, we explore fidelity dynamics of quantum teleportation with the two atoms prepared initially in the product state $|e_1\rangle \otimes |g_2\rangle$. The density matrix $\rho(t)$ can be obtained directly from Eqs. (9) and (10) with the initial condition $\rho_{11}(0) = 1$, from which one can obtain the concurrence analytically as $C = e^{-\gamma t}[(e^{-\gamma_{12}t} - e^{\gamma_{12}t})^2 + 4\sin^2(2\Omega_{12}t)]^{1/2}/2$. Clearly, for very small interatomic separations $r_{12}$ the dipole-dipole interaction potential $\Omega_{12}$ becomes very strong and thus one can entangle the two atoms with considerable amount of entanglement (e.g., when $r_{12} = \lambda/20$ we have $C_{\max} \simeq 0.9688$ at the critical time $\gamma t_c \simeq 0.032$, and $C_{\max}$ may be further increased by decreasing $r_{12}$). Since entanglement of $\rho(t)$ is the prerequisite for teleportation, now we see whether this entanglement can be used to teleport the one-qubit state with $F[\rho(t)] > 2/3$. The four parameters defined in Eq. (5) can be derived analytically as

$$\chi_{0,3} = \frac{1}{2} - \frac{1}{4}e^{-(\gamma+\gamma_{12})t} - \frac{1}{4}e^{-(\gamma-\gamma_{12})t}, \quad \chi_{1,2} = \frac{1}{2}e^{-(\gamma\pm\gamma_{12})t}. \tag{17}$$

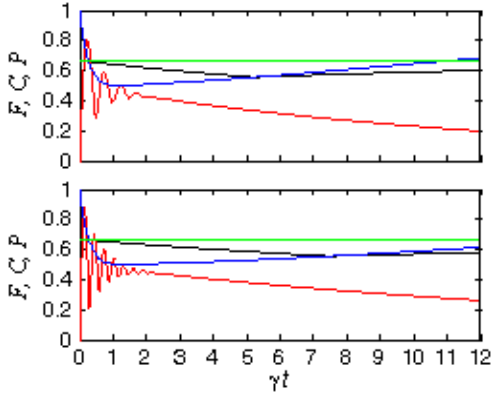

**Figure 6.** Average fidelity of teleportation (black curves), concurrence (red curves) and purity (blue curves) of the atomic state $\rho(t)$ versus the rescaled time $\gamma t$ with $\rho(0) = |e_1 g_2\rangle\langle e_1 g_2|$ and $\hat{\mu} \perp \hat{r}_{12}$. The other parameters for the plots are $r_{12} = \lambda/10$ (the top panel) and $r_{12} = \lambda/12$ (the bottom panel). Moreover, the green horizontal lines at $F = 2/3$ show the highest fidelity for classical transmission of a quantum state.

From Eq. (17) one can note that the dipole-dipole interaction potential $\Omega_{12}$ though appearing in the density operator $\rho(t)$ does not contribute to the average fidelity $F[\rho(t)]$. Moreover, the relative magnitudes of $\chi_m$ ($m = 0, 1, 2, 3$) depend on the parameters $r_{12}$ and $t$ involved and their maximum cannot be written compactly unless the parameter values are known. Thus instead of performing the analytical analysis, we presented in figure 6 the time dependence of the average fidelity $F[\rho(t)]$, as well as the concurrence $C$ and the purity $P$ of $\rho(t)$ with interatomic separations $r_{12} = \lambda/10$ and $r_{12} = \lambda/12$, respectively. From Eq. (2) one can obtain numerically that when $r_{12} < 0.4366\lambda$ we always have the collective damping rate $\gamma_{12} > 0$, thus the fully entangled fraction can be expressed as $\mathcal{F}[\rho(t)] = \chi_2$ when the inequality $e^{-\gamma_{12}t} + 3e^{\gamma_{12}t} > 2e^{-\gamma t}$ holds, for which the one-to-one correspondence between Alice's measurement basis and Bob's recovery transformation should be $\Pi_{A1}^{0,1,2,3} \mapsto \sigma^{2,3,0,1}$ for achieving the maximal average fidelity $F[\rho(t)] = [e^{-(\gamma-\gamma_{12})t} + 1]/3$. Clearly, although one can entangle the two atoms greatly via the spontaneous emission and dipole-dipole interactions of the system, $F[\rho(t)]$ still cannot exceed the classical limiting value of 2/3 [17] since $\gamma_{12} < \gamma$ and thus the exponential term $e^{-(\gamma-\gamma_{12})t}$ is always smaller than unity. This confirms again that entanglement is the prerequisite but not the only essential quantity for predicting the fidelity. Similarly, when $e^{-\gamma_{12}t} + 3e^{\gamma_{12}t} < 2e^{-\gamma t}$ one can obtain $\mathcal{F}[\rho(t)] = \chi_{0,3}$, and thus the one-to-one correspondence becomes $\Pi_{A1}^{0,1,2,3} \mapsto \sigma^{0,1,2,3}$ or $\Pi_{A1}^{0,1,2,3} \mapsto \sigma^{3,2,1,0}$, which yields $F[\rho(t)] = 2/3 - e^{-(\gamma+\gamma_{12})t}/6 - e^{-(\gamma-\gamma_{12})t}/6$. As can be seen clearly from figure 6, $F[\rho(t)]$ though increases with increasing time $\gamma t$ still cannot exceed 2/3 for the obvious facts that both the exponential terms $e^{-(\gamma+\gamma_{12})t}$ and $e^{-(\gamma-\gamma_{12})t}$ are always larger than zero, no matter how one controls the interatomic separations $r_{12}$ or the decay time $\gamma t$.

Since in all the above discussions we only considered the case of $\hat{\mu} \cdot \hat{r}_{12} = 0$, i.e., the atomic dipole moments are polarized in the direction perpendicular to the interatomic axis, it is natural for us to wonder whether the average fidelity can be enhanced by tuning the directions of $\hat{\mu}$. In fact, this issue can be analyzed similarly as the previous section. From equation (2) one can obtain numerically that if the two atoms are separated by small distances, the collective damping rate $\gamma_{12}$ will be increased slightly when the separation angle between $\hat{\mu}$ and $\hat{r}_{12}$ decreases from $\pi/2$ to $0$, thus from equations (12) and (15) one can see that in the region of $(\gamma + \gamma_{12})t < \ln 3$, the average fidelity $F[\rho(t)]$ will be decreased to some extent for initial atomic state $|\Psi^1\rangle$, while the reverse situation occurs for $F[\rho(t)]$ with initial atomic state $|\Psi^2\rangle$ in the region of $(\gamma - \gamma_{12})t < \ln 3$. If we continue increasing the interatomic separations, $\gamma_{12}$ will become smaller or larger than that with $\hat{\mu} \cdot \hat{r}_{12} = 0$, and the same (reverse) behavior happens for $F[\rho(t)]$ with initial $|\Psi^2\rangle$ ($|\Psi^1\rangle$). Moreover, for initial atomic states $|\Psi^{0,3}\rangle$, the average fidelity $F[\rho(t)]$ will be increased with small interatomic separations and decreased with large interatomic separations. Finally, From Eq. (17) one can see that no matter how one tuning the directions of the atomic dipole moments, the average fidelity $F[\rho(t)]$ cannot exceed the classical limiting value $2/3$ for $\chi_m$ ($m = 0,1,2,3$) is always smaller than $1/2$.

## 4. Summary and discussion

In summary, we have investigated quantum teleportation of the one-qubit state by using two spatially separated two-level atoms as the quantum channel. We initially presented formula for evaluating quality of the teleported state under general conditions, i.e., the fidelity averaged over all possible pure input states on the Bloch sphere and all possible Alice's measurement outcomes. Then based on this formula, we explored fidelity dynamics of teleportation with $\rho(t)$ generated from different initial atomic states, including the maximally entangled Bell states and the product state, and for all of them, we gave explicit operations performed by Alice and Bob for achieving the maximal average fidelity.

We have revealed that for short decay times, the state can be teleported nonclassically over a distance larger than a critical value if the two atoms are prepared initially in the symmetric state $|\Psi^1\rangle$, and over arbitrary long distances if they are prepared initially in the antisymmetric state $|\Psi^2\rangle$. Increasing the decay time will shorten the region of interatomic separations for nonclassical teleportation, and for long decay times the state will cannot be teleported with $F[\rho(t)] > 2/3$ for $|\Psi^1\rangle$, and can only be teleported over finite distances for $|\Psi^2\rangle$. We also demonstrated that for fixed decay time $\gamma t$, the average fidelity attains its maximum when the two atoms are separated by distance of about $0.6737$ times of the resonant wavelength and the dipole transition moments are polarized in the direction perpendicular to the interatomic axis for initial state $|\Psi^1\rangle$, while for $|\Psi^2\rangle$, $F[\rho(t)]$ increases with decreasing interatomic separations and attains its maximum for the small sample model. Moreover, if the two atoms are prepared initially in the states $|\Psi^{0,3}\rangle$, our results show that the average fidelity exhibits completely the same dynamical behaviors, which are significantly influenced by the transition frequency of the atoms. $F[\rho(t)]$ attains its maximum for the small sample model and oscillates rapidly with increasing interatomic separations.

Besides the four maximally entangled Bell states, we also presented results of average fidelity with the system prepared initially in a product state $|e_1\rangle \otimes |g_2\rangle$, i.e., the case with only the first atom of the channel excited. Analytical and numerical analysis show that although from this initial separable state one can entangle the two atoms via the dissipative process of spontaneous emission with considerable amount of transient entanglement, it cannot be adopted to serve as a quantum channel for teleportation because we always have the average fidelity $F[\rho(t)] < 2/3$.

We would also like to emphasize that although entanglement of the channel state $\rho(t)$ is the prerequisite for implementing the teleportation protocol, our results presented in this work clearly shows that it is not the only essential quantity for predicting the fidelity. In a recent work we have shown that for some specific channels the purity of $\rho(t)$ is also important in determine quality of teleportation [16]. However, we found here the situations (i.e., the two atoms prepared initially in the states $|\Psi^{0,3}\rangle$) in which the average fidelity increases while both the entanglement and the purity of $\rho(t)$ decrease. Although this phenomenon can be interpreted mathematically by the presence of the phase factors $e^{\pm 2i\omega_0 t}$ appearing in $\rho_{41,14}(t)$, it remains still an open question which quantities ultimately determine the teleportation fidelity for general cases? We hope our results will stimulate more future works which may give satisfactory physical explanations of this seemingly counterintuitive phenomenon.

## Acknowledgments

This work was supported in part by the NSF of Shaanxi Province under grant Nos. 2010JM1011 and 2009JQ8006, the Specialized Research Program of Education Department of Shaanxi Provincial Government under grant Nos. 2010JK843 and 2010JK828, and the Youth Foundation of XUPT under Grant No. ZL2010-32.

[26] Dicke R H 1954 Phys. Rev. 93 99